\documentstyle[12pt,epsf]{article}
\newcommand{\beq}{\begin{equation}}
\newcommand{\eeq}{\end{equation}}
\newcommand{\ba}{\begin{array}}
\newcommand{\ea}{\end{array}}
\newcommand{\beqa}{\begin{eqnarray}}
\newcommand{\eeqa}{\end{eqnarray}}
\newcommand{\dis}{\displaystyle}
\newcommand{\cL}{{\cal L}}

\newcommand{\cO}{{\cal O}}
\newcommand{\IcA}{{\Im m\cal A}}
\newcommand{\RcA}{{\Re e\cal A}}

\newcommand{\no}{\nonumber}
\newcommand{\lsim}{\stackrel{<}{_\sim}}
\newcommand{\gsim}{\stackrel{>}{_\sim}}

\newcommand{\Real}{\Re e}

\newcommand{\eps}{\epsilon}

\newcommand{\PL}[3]{{Phys. Lett.}       {\bf #1} {(19#2)} {#3}}

\newcommand{\PRL}[3]{{Phys. Rev. Lett.} {\bf #1} {(19#2)} {#3}}
\newcommand{\PR}[3]{{Phys. Rev.}        {\bf #1} {(19#2)} {#3}}
\newcommand{\NP}[3]{{Nucl. Phys.}       {\bf #1} {(19#2)} {#3}}

\begin{document}    
\thispagestyle{empty}
\setcounter{page}{0}
\begin{flushright}
INFNNA-IV-97/40\\
DSFNA-IV-97/40\\
{\tt hep-ph/9708326}\\
August 1997

\end{flushright}
\vspace*{1cm}
\centerline{\Large\bf Can we extract short--distance }
\vspace*{0.3cm}
\centerline{\Large\bf information from $B(K_L \to \mu^+ \mu^-)$ ?}
\vspace*{1.5cm}
\centerline{{\sc G. D'Ambrosio${}^a$, G. Isidori${}^{b}$ 
\mbox{\rm and} J. Portol\'es${}^a$} }
\bigskip
\vspace*{0.5cm}
\centerline{\sl ${}^a$INFN, Sezione di Napoli,
                Dip. di Scienze
                Fisiche, Univ. di Napoli,  
                I-80125 Napoli, Italy}
\centerline{\sl ${}^b$INFN, Laboratori Nazionali di Frascati, 
                P.O. Box 13, I-00044 Frascati, Italy}
\vspace*{1.5cm}
\centerline{\bf Abstract}
\vspace*{0.2cm}
\noindent 
A new analysis of  the  
long--distance two--photon dispersive amplitude of 
$K_L \to \mu^+ \mu^-$ is presented. We introduce a phenomenological
parameterization of the $K_L \to\gamma^*\gamma^*$ form factor, 
constrained at low energies by $K_L\to\gamma \ell^+ \ell^-( \ell=e,\mu)$
data
and at high energies by perturbative QCD. 
Using this form factor we provide a reliable estimate of
magnitude and relative uncertainty of the two--photon dispersive 
contribution to $K_L \to \mu^+ \mu^-$. We finally discuss
the implications of this analysis 
for the extraction of short--distance information from  
$B(K_L \to \mu^+ \mu^-)$.
\vfill
\newpage
\pagenumbering{arabic}

\section{Introduction}   
Historically the $K_L \to \mu^+ \mu^-$ decay
provided a very important tool for understanding the flavour structure 
of electroweak interactions \cite{GIM,GL} and nowadays 
it still represents an interesting window on short--distance dynamics.
The amplitude of this process
can be conveniently decomposed into two distinct parts: a 
long--distance contribution generated by the two--photon intermediate
state
(Fig.~1a) and a short--distance part that, within the Standard Model, is 
due to $W$ and $Z$ exchange (Fig.~1b). The latter turns out to be
dominated by 
the top quark and it is known to the next--to--leading order in QCD
\cite{BB}. 
If we were able to disentangle this contribution from the
measured $K_L \to \mu^+ \mu^-$ branching ratio we could extract 
interesting information on the  Cabibbo--Kobayashi--Maskawa (CKM) matrix 
element $V_{td}$ \cite{BNL}. Furthermore, a model--independent 
determination of the short--distance amplitude could be useful to 
put constraints on possible Standard Model extensions \cite{Nir}. 

To fully exploit 
the potential of $K_L \to \mu^+ \mu^-$ in probing 
short--distance dynamics, it is necessary to have a reliable control on 
its long--distance amplitude. However, the dispersive contribution 
generated by the two--photon intermediate state cannot be 
calculated in a model--independent way and it is subject to various 
uncertainties \cite{VS,BMS1,DGE,BG,KO,EKP}. The purpose of this paper
is to 
re--analyse this contribution, using all available information 
on the $K_L \to\gamma^*\gamma^*$ transition and trying to evaluate
the error due to the model dependent assumptions. 
We will introduce a new low--energy parameterization of the 
$K_L \to\gamma^*\gamma^*$ form factor in terms of two parameters 
$\alpha$ and $\beta$ measurable from $K_L\to\gamma \ell^+ \ell^-
( \ell=e,\mu)$ and
$K_L\to e^+ e^- \mu^+\mu^-$. Moreover, we will discuss the 
matching of this approach with the behaviour of the form factor
in perturbative QCD. Finally, using our estimate of  the
two--photon dispersive contribution,  we will 
derive new bounds on the CKM parameter  $\rho$ \cite{WolfCKM}
and on possible new--physics flavour--changing couplings.

The plan of the paper is as follows. In Section 2 we
briefly discuss the general decomposition of the $K_L \to \mu^+ \mu^-$ 
branching ratio and the main formulae for the bounds on short--distance 
parameters. In Section 3 we introduce our low--energy parameterization 
of the $K_L \to\gamma^*\gamma^*$ form
factor, we discuss the determination of $\alpha$ and
$\beta$ and the matching with the QCD calculation.
Finally, in Section 4, we analyse the numerical results.

\begin{figure}[t]
    \begin{center}
       \setlength{\unitlength}{1truecm}
       \begin{picture}(4.0,2.5)
       \epsfxsize 4.  true  cm
       \epsfysize 2.5  true cm
       \epsffile{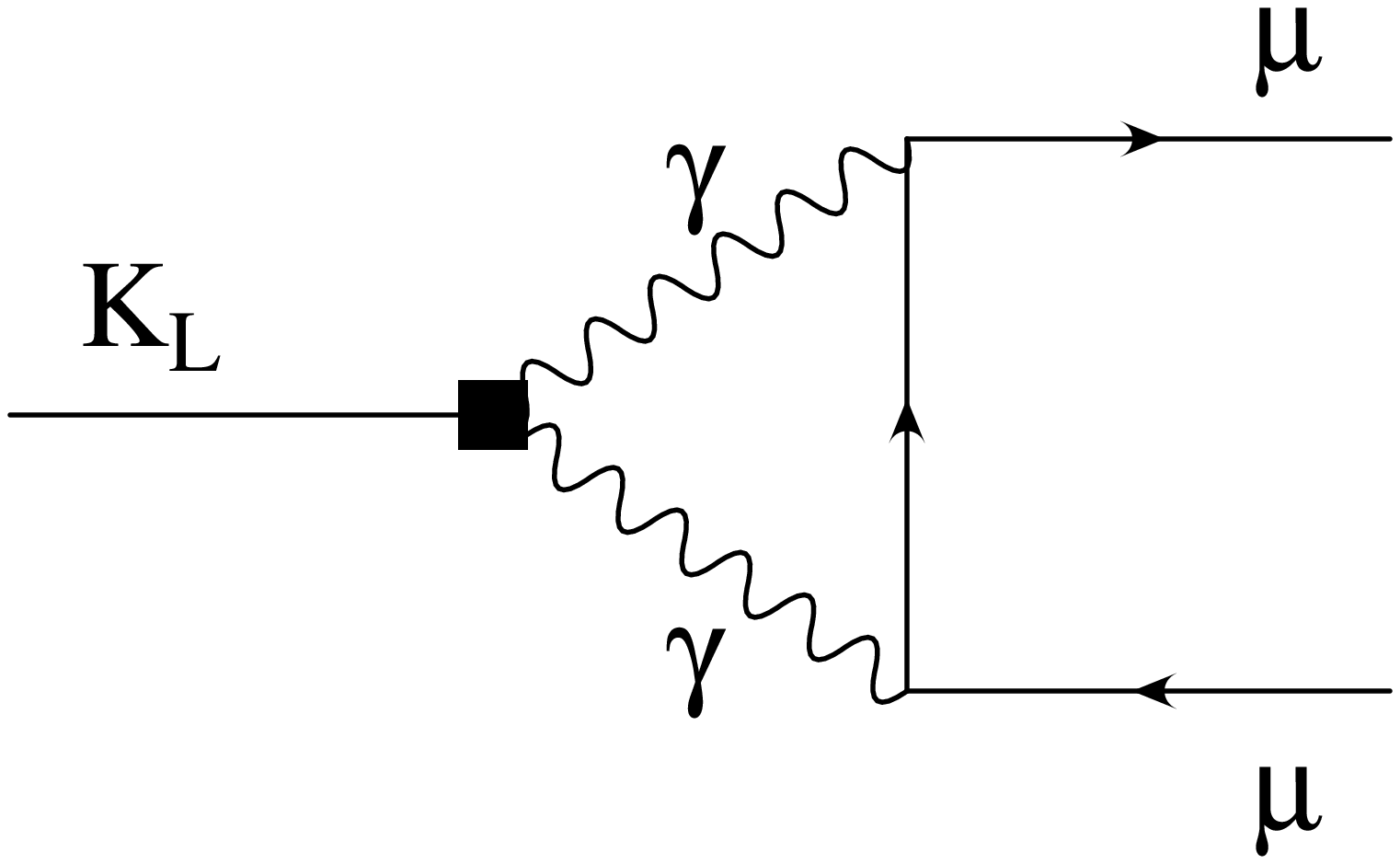}
       \end{picture}
       \centerline{a)}
       \begin{picture}(13.5,2.5)
       \epsfxsize 13.5  true  cm
       \epsfysize 2.5  true cm
       \epsffile{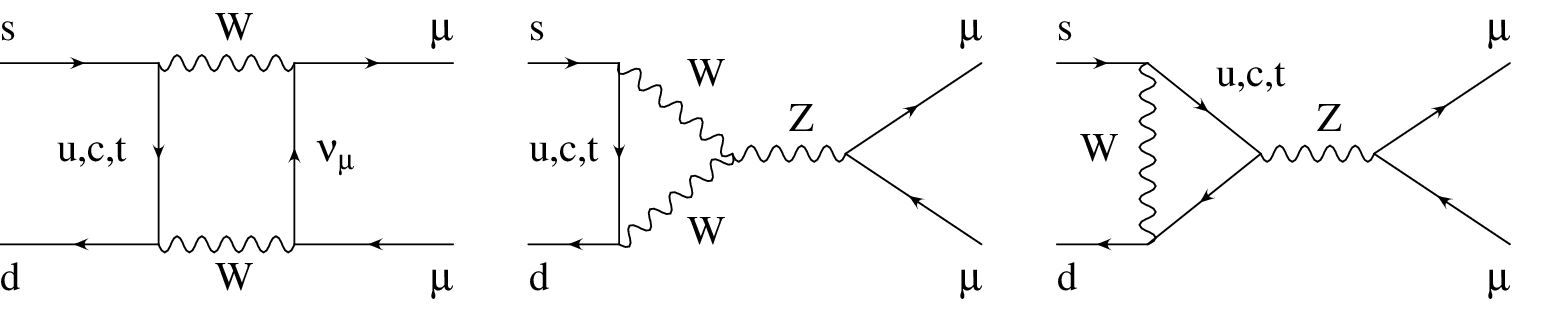}
       \end{picture}       
      \centerline{b)}
    \end{center}
    \caption{Long--distance (a) and lowest--order 
 short--distance (b) contributions to 
                 $K_L \to \mu^+\mu^-$.  }
    \protect\label{fig:1}
\end{figure}
%

\section{Decomposition of $B(K_L \to \mu^+ \mu^-)$}
The $K_L \to \mu^+ \mu^-$  branching ratio
can be generally decomposed in the following way
\beq
B(K_L \to \mu^+ \mu^-) = |\RcA|^2 + |\IcA|^2~,
\eeq
where $\RcA$ denotes the dispersive contribution and 
$\IcA$  the absorptive one. The former can be rewritten as
\beq
\RcA =  \RcA_{long} + \RcA_{short}~,
\label{eq:2}
\eeq
whereas the latter can be  
determined in a model independent way from the 
$K_L\to\gamma\gamma$ branching ratio\footnote{~In principle 
the absorptive amplitude also receives contributions  
from intermediate states other than two--photons, 
like the two--pions one, but these are completely negligible
\protect\cite{deRafael}.}
\beq
|\IcA|^2 =  \frac{\alpha_{em}^2 m_\mu^2}{2 m_K^2\beta_\mu }
\left[\ln \frac{1-\beta_\mu}{1+\beta_\mu} \right]^2 
B(K_L \to \gamma\gamma)~,\qquad\quad
\beta_{\mu} = \sqrt{1-\frac{4m_\mu^2}{m_K^2}}~.
\label{eq:immgg}
\eeq
The recent measurement of $B(K_L \to \mu^+ \mu^-)$ \cite{BNL} is 
almost saturated by the value of $|\IcA|^2$, leaving 
a very small room for the dispersive contribution \cite{BNL} 
\beqa
|\RcA_{exp}|^2 &=& B(K_L \to \mu^+ \mu^-) - |\IcA|^2 
= (-1.0 \pm 3.7)\times10^{-10}  \qquad \mbox{or} \no \\
|\RcA_{exp}|^2 &<&  5.6 \times 10^{-10} \qquad (90\%~\mbox{C.L.})~. 
\eeqa

Within the Standard Model the   
NLO short--distance amplitude can be written as \cite{BB,Buras}
\beq
|\RcA_{w}|^2 = 0.9\times 10^{-9} (1.2 - \bar\rho)^2 \left[ 
\frac{ {\overline m_t}(m_t) }{170~\mbox{GeV}} \right ]^{3.1} 
\left[ \frac{|V_{cb}|}{0.040}\right]^4~,
\eeq
where $\bar\rho=\rho(1-\lambda/2)$ \cite{BurasCKM} and 
$\rho$, $\lambda$ are the usual Wolfenstein parameters \cite{WolfCKM}. 
Using this result we can write
\beq
\bar\rho = 1.2 -
\left| \frac{|\RcA_{exp}| \pm |\RcA_{long}| }{3 \times 10^{-5} } \right| 
\left[ \frac{ {\overline m_t}(m_t) }{170~\mbox{GeV}} \right ]^{-1.55} 
\left[ \frac{|V_{cb}|}{0.040}\right]^{-2}~,
\eeq
where the sign inside the modulus is positive if $\RcA_{w}$ and 
$\RcA_{long}$ interfere destructively and 
$|\RcA_{w}|>|\RcA_{long}|$. In principle the above equation could 
be used to put both a lower and an upper bound on $\bar\rho$.
However  $|\RcA_{exp}|$ is compatible with zero and, 
as we will show in the following, the same is true for 
$|\RcA_{long}|$. Thus the upper bound on $\bar\rho$ is 
useless since it is above unity. On the other hand, independently of the
interference sign between $\RcA_{w}$ and $\RcA_{long}$, we can 
derive a possibly meaningful lower bound on $\bar\rho$
\beq
\bar\rho > 1.2 - \mbox{max}
\left\{ \frac{|\RcA_{exp}| + |\RcA_{long}| }{3\times 10^{-5} } 
\left[ \frac{ {\overline m_t}(m_t) }{170~\mbox{GeV}} \right ]^{-1.55} 
\left[ \frac{|V_{cb}|}{0.040}\right]^{-2}\right\} ~.
\label{eq:rbound}
\eeq

Beyond the Standard Model we can parameterize new--physics 
contributions as in \cite{Nir},  
introducing a flavour--changing $Zds$ coupling at the tree level.
Using the Lagrangian 
\beq
\cL_{NP}^Z = {g\over 2 \cos\theta_w } \sum_{i\not=j}
 U_{ij} {\bar d}_L^i \gamma^\mu d_L^j  Z_\mu~,
\eeq
we obtain $|\RcA_{NP}| = 3.7 \,|\Real U_{ds}|$. Then, assuming
$\RcA_{short}=\RcA_{NP}+\RcA_{w}$, the most conservative
bound on $|\Real U_{ds}|$ is given by
\beq
|\Real U_{ds}| < 0.27 \,\mbox{max} 
\left\{ |\RcA_{exp}| + |\RcA_{long}|+ |\RcA_{w}| \right\}~.
\label{eq:ubound}
\eeq

\section{The $K_L\to\gamma^*\gamma^*$ form factor and $\RcA_{long}$ }
The necessary ingredient for the evaluation of $\RcA_{long}$ is the 
construction of a suitable $K_L \to\gamma^*\gamma^*$ amplitude. Assuming
$CP$ conservation, gauge and Lorentz invariance implies the 
following general decomposition \cite{EPR}
\beq
A(K_L \to \gamma^*(q_1,\eps_1)\gamma^*(q_2,\eps_2))
= i \varepsilon_{\mu\nu\rho\sigma} \, 
\eps_1^\mu\eps_2^\nu q_1^\rho q_2^\sigma \, F(q_1^2,q_2^2)~,
\label{eq:defF}
\eeq
where $F(q_1^2,q_2^2)$ is a symmetric function under the interchange 
of $q_1^2$ and $q_2^2$, and 
$|F(0,0)|$ can be determined 
by the $K_L\to\gamma\gamma$ width \cite{PDG}
\beq
|F(0,0)|=\left[\frac{64\pi\Gamma(K_L\to\gamma\gamma)}{m_K^3}\right]^{1/2}
= (3.51\pm 0.05)\times 10^{-9} \mbox{GeV}^{-1}~.
\eeq
Using (\ref{eq:defF}) we obtain
\beq
|\RcA_{long}|^2 =  \frac{2\alpha_{em}^2m_\mu^2\beta_\mu}{\pi^2m_K^2} \, 
B(K_L\to\gamma\gamma) \, |\Real R(m^2_K)|^2~,
\label{eq:ral11}
\eeq
where \cite{ABM}
\beq
R(q^2)=\frac{2i}{\pi^2 q^2 }\int\mbox{d}^4\ell~ 
\frac{q^2\ell^2 -(q\cdot \ell)^2}{\ell^2(\ell-q)^2[(\ell-p)^2-m_\mu^2]}
\frac{F(\ell^2,(\ell-p)^2)}{F(0,0)}
\label{eq:Rdef}
\eeq 
and $p^2=m_\mu^2$. 

The structure of the $K_L \to\gamma^*\gamma^*$
form factor has already been discussed and parameterized in 
different ways in the literature \cite{VS,BMS1,BG,KO,EKP}. 
However, all the existing analyses use model
dependent assumptions and thus suffer 
from uncontrolled theoretical
uncertainties. In order to be as model independent as 
possible and to evaluate the size of the theoretical errors,
we propose the following low--energy parameterization 
\beq
f(q_1^2,q_2^2) = \frac{F(q_1^2,q_2^2)}{F(0,0)} = 
1+ \alpha\left( \frac{q_1^2}{q_1^2 - m_V^2} +
\frac{q_2^2}{q_2^2 - m_V^2} \right) + \beta \frac{q_1^2 q_2^2}{ 
(q_1^2 - m_V^2)(q_2^2 - m_V^2) }~,
\label{eq:fdef}
\eeq
where $\alpha$ and $\beta$ are arbitrary real parameters and 
$m_V$ is conventionally chosen to be the $\rho$ mass. The above 
expression has at least three interesting features:
\begin{enumerate}
\item{ It is the most general parameterization compatible with 
the chiral expansion of the 
$K_L\to\gamma^*\gamma^*$ amplitude up to $O(p^6)$ \cite{EPR,DEIN}.  }
\item{ It includes the poles of the lowest vector meson 
resonances with arbitrary residues.}
\item{ The parameters $\alpha$ and $\beta$, expected to
be $\cO(1)$  by naive dimensional chiral power
counting, are in principle directly accessible by experiments in
$K_L\to\gamma \ell^+ \ell^-(\ell=e,\mu)$ and
$K_L\to e^+ e^- \mu^+\mu^-$. }
\end{enumerate}

Clearly the expression (\ref{eq:fdef}) cannot be considered
correct for arbitrary values of $q^2_1$ and $q^2_2$. To be more general
we should consider $\alpha$ and $\beta$ as $q^2$--dependent 
couplings. However for the purposes of this first analysis (that should
be improved in the future along these suggestions) we believe 
it is reasonable to treat
$\alpha$ and $\beta$ as constants up to 
$q^2_1 \sim q^2_2 \sim 1$ GeV$^2$. 
Moreover, being just a phenomenological description,
we do not expect the form factor (\ref{eq:fdef}) to produce
a finite result for the $K_L\to \mu^+\mu^-$ amplitude. Indeed,
using (\ref{eq:Rdef}) and (\ref{eq:fdef}) we obtain
\beqa
\Real R(m^2_K) &=& -3 \big[ \ln(\Lambda/m_0)
+2\alpha\ln(\Lambda/m_\alpha)
+\beta\ln(\Lambda/m_\beta) \big] \no \\
&=&   -3 \big[ \ln(m_\beta/m_0) +2\alpha\ln(m_\beta/m_\alpha) \big]
-3 (1+2\alpha+\beta) \ln(\Lambda/m_\beta)~, 
\label{eq:Rint}
\eeqa
where 
\beq
m_0 = 140~\mbox{MeV}~, \qquad m_\alpha = 452~\mbox{MeV}~,
\qquad m_\beta = 806~\mbox{MeV}~,
\eeq
and $\Lambda$ is an ultraviolet cutoff. 
As one could expect from (\ref{eq:Rdef}), the cutoff sensitivity of 
(\ref{eq:Rint}) is determined by the value of the combination 
$(1+2\alpha+\beta)$. Indeed, for large values of the loop--momentum, 
the integrand in (\ref{eq:Rdef}) is proportional to 
\beq
f(\ell^2,\ell^2) \stackrel{\ell^2\gg m_V^2}{\longrightarrow } 
1 + 2\alpha+\beta~.
\label{eq:flim}
\eeq

The following subsections are devoted to the
determination of $\alpha$ and $\beta$. At first 
we shall analyse the experimental 
information coming from $K_L\to \gamma \ell^+ \ell^-$ and 
$K_L\to \mu^+\mu^-e^+e^-$. Then we will 
constraint the value of $(1+2\alpha+\beta)$ by studying the behaviour 
of $f(q^2,q^2)$ at large $q^2$ in the framework 
of perturbative QCD. Finally we will discuss the consistency of
the previous findings with a model--dependent  
determination of $\alpha$ and $\beta$ within the 
approach proposed in \cite{DP}.

\subsection{Experimental determination of $\alpha$ and $\beta$}  
As anticipated, $\alpha$ and $\beta$ are in principle  
accessible by experiments in the decays $K_L\to \gamma \ell^+ \ell^-$ and 
$K_L\to \mu^+\mu^-e^+e^-$, dominated by the
$K_L \rightarrow \gamma \gamma^*$ and $K_L \rightarrow \gamma^*
\gamma^*$ form factors respectively.
The differential decay rates of $K_L\to \gamma \ell^+ \ell^-$ and 
$K_L\to \mu^+\mu^-e^+e^-$, normalized to $\Gamma_L^{\gamma \gamma} 
\equiv \Gamma(K_L \rightarrow \gamma \gamma)$, are given by
\beqa
\frac{1}{\Gamma_L^{\gamma\gamma}}
\frac{\mbox{d}\Gamma_L^{\ell^+ \ell^-\gamma} }{\mbox{d}q^2} &=& 
\frac{2}{ q^2 } \left(\frac{\alpha_{em}}{3\pi}\right)
|f(q^2,0)|^2 \, \, \lambda^{3/2} \left(1,{q^2\over m_K^2},0 \right) 
\, G_{\ell}(q^2)~, 
\eeqa
\beqa
\frac{1}{\Gamma_L^{\gamma\gamma}}
\frac{\mbox{d} \Gamma_L^{\mu^+\mu^-e^+e^-} }{\mbox{d}q^2_e
\mbox{d}q^2_\mu } &=& \frac{2}{ q_e^2q_\mu^2 }  
\left(\frac{\alpha_{em}}{3\pi}\right)^2 |f(q^2_e,q^2_\mu)|^2 
 \lambda^{3/2} \left(1,{q^2_e\over m_K^2},{q^2_\mu\over m_K^2} \right)
 G_e(q^2_e)  G_\mu(q^2_\mu),
\eeqa
where
\beq
\lambda(a,b,c) = a^2+b^2+c^2 -2(ab+bc+ac)
\eeq
and
\beq
G_{\ell}(q^2) = \left(1-\frac{4m_{\ell}^2}{q^2}\right)^{1/2}
\left(1+\frac{2m_{\ell}^2}{q^2}\right)~.
\eeq  

Present  data on both $K_L\to \gamma e^+e^-$ \cite{Kleeg1,Kleeg2}
and $K_L\to \gamma \mu^+\mu^-$ \cite{Klmmg} let us extract 
useful information about the $q^2$ dependence of $f(q^2,0)$. 
The experimental results have been analysed up to now  
assuming only the form factor proposed by 
Bergstr\"om, Mass\'{o} and Singer (BMS model) \cite{BMS1}. 
The latter depends on one unknown parameter $\alpha_K^*$ and, 
expanding in powers of $q^2/m_\rho^2$, can be written as  
\beq
f(q^2,0)_{BMS} \simeq  1 + (1-3.1\alpha_K^*)\frac{q^2}{m^2_\rho}
 +\cO\left( (q^2)^2/m^4_\rho)\right)~.
\label{eq:BMSff}
\eeq
The fitted values of  $\alpha_K^*$ are given by
\beqa
\alpha_K^* &=& -0.280 \pm 0.083~^{+0.054}_{-0.034}~\cite{Kleeg1}~, \no\\
\alpha_K^* &=& -0.28  \pm 0.13~\cite{Kleeg2}~, \\
\alpha_K^* &=& -0.028~^{+0.115}_{-0.111}~\cite{Klmmg}~, \no 
\eeqa
and the corresponding weighted average is
\beq
\alpha_K^*=-0.204\pm 0.062~.
\label{eq:alphaK}
\eeq
Comparing the BMS form factor (\ref{eq:BMSff}) 
with the one proposed in (\ref{eq:fdef}), we obtain the following
relation
\beq
\alpha= -1 + (3.1 \pm 0.5) \alpha_K^*~,
\eeq
where the error is due to the different  
quadratic dependence on $q^2/m_\rho^2$. Then, using (\ref{eq:alphaK}) we
find
\beq
\alpha= -1.63 \pm 0.22~. 
\label{eq:alphaval}
\eeq
As already pointed out in \cite{DP}, it must be stressed that 
an improved determination of $\alpha$ would be possible if
present data were not analysed assuming only the BMS model. 
Furthermore, an experimental determination of the quadratic 
term in the expansion of $f(q^2,0)$  would be extremely useful 
in order to perform a consistency check of our approach.
\par
Contrary to $\alpha$, the experimental determination
of $\beta$ is much difficult. 
In principle the $K_L \rightarrow e^+ e^- \mu^+ \mu^-$ rate 
should be sensitive,  in the region where 
both di--lepton pairs have a large invariant mass, 
to the higher structure in momenta
carried by the $\beta$ component of the form factor. However, 
the real sensitivity of this process to $\beta$ is rather small. 
Thus, even though the first evidence for
$K_L \rightarrow e^+ e^- \mu^+ \mu^-$
has been recently reported \cite{Klmmee}, it is unlikely 
that $\beta$ will be measured with a reasonable accuracy 
in the short term.

\subsection{Perturbative evaluation of $f(q^2,q^2)$}
In the limit $q_1^2=q_2^2=q^2 \gg m_K^2$ we can simply evaluate
the form factor within perturbative QCD. 
At the lowest order in $\alpha_s$, the only diagrams that contribute to 
$f(q^2,q^2)$ are those shown in Fig.~2 \cite{VS}.
Neglecting masses and momenta of the external quarks, as well as 
the contribution of the top quark inside the loop
(suppressed by CKM factors \cite{EKP}), the result can be written as
\beq
f^{QCD}(q^2,q^2) = N_F \left[ g_u\left({q^2\over 4m^2_u}\right)
                 - g_c\left({q^2\over 4m^2_c}\right) \right]~,
\label{eq:fQCD}
\eeq
where
\beq
g_q(r)= ~-~ r\frac{\mbox{d}}{\mbox{d}r} J(r)~+~
 \left[ {1+2r\over 6r} J(r) ~+~ {1\over 3}
\ln{M^2_W\over m^2_q} \right]
\label{eq:gdef}
\eeq
and 
\beq
J(r)=\dis\left\{ \ba{ll}
\dis -2\sqrt{1/r-1}\arctan\sqrt{r/(1-r)} 
   +2 &  \qquad\qquad 0< r < 1~, \\ 
& \\
 \dis \sqrt{1-1/r}\left(
\ln\frac{1-\sqrt{1-1/r}}{1+\sqrt{1-1/r}}+i\pi\right)+2 &
\qquad\qquad r > 1~. \end{array} \right. 
\eeq
The normalization factor of (\ref{eq:fQCD}) is given by 
\beq
|N_F| =  \frac{16}{9}  \frac{\lambda G_F F_\pi \alpha_{em} 
}{|F(0,0)| \pi\sqrt{2}} \simeq 0.20~,
\eeq
where $\lambda$ denotes the sine of the Cabibbo angle \cite{WolfCKM}
and $F_\pi\simeq 93$ MeV the pion decay constant.
The first term in (\ref{eq:gdef})
is the contribution of the diagram in Fig.~2a, whereas the second one is
originated by the graphs in Fig.~2b--c. We have neglected 
all the contributions independent of quark masses that cancel via
the GIM mechanism and, whenever possible, we have considered the limit
$M_W \to \infty$ (this is always possible except for the 
$\ln(M^2_W/m^2_q)$ term originated by the diagrams in Fig.~2b--c).

\begin{figure}[t]
    \begin{center}
      \setlength{\unitlength}{1truecm}
       \begin{picture}(13.5,2.5)
       \epsfxsize 13.5  true  cm
       \epsfysize 2.5  true cm
       \epsffile{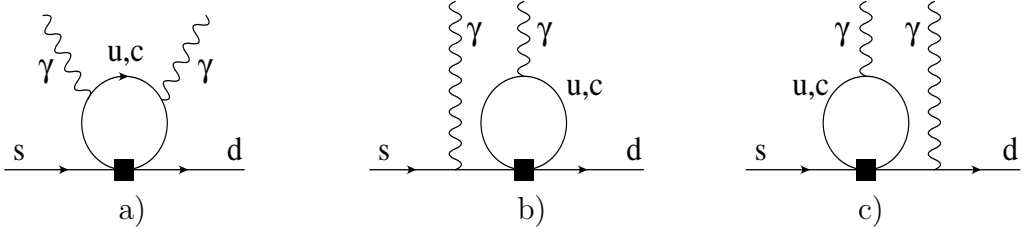}
       \end{picture} \\ 
    \hglue -0.3 true cm a) \hglue 4.8 true cm b) \hglue 4.0 true cm  c) 
    \end{center}
    \caption{Lowest--order quark diagrams that contribute to the 
$K_L \to \gamma^* \gamma^*$ transition (every diagram is also understood 
with the corresponding photons crossed). }
    \protect\label{fig:2}
\end{figure}

From the above equations it follows
\beq
|\Real f^{QCD}(q^2,q^2)| = |N_F| \left\{ \ba{ll}
 \cO(m_c^2/q^2) &  q^2 \gg 4m_c^2~, \\
\dis{14\over 9} + {1\over 3} \ln{m_c^2\over q^2} \qquad\qquad
 & 4m_u^2 \ll q^2 \ll 4m_c^2~. \ea \right.
\label{eq:fqcdapp}
\eeq
Using this approximate expression in (\ref{eq:Rdef}) and keeping 
in the final result only the dominant $\ln(m^2_c/m^2_u)$ terms, 
leads to the approximate formula of Voloshin and Shabalin
for $\RcA_{long}$ \cite{VS}.
This result indicates that the long--distance dispersive 
amplitude of $K_L \to \mu^+\mu^-$ is very small, 
however it cannot be trusted in detail since the low 
$q^2$ limit of $f^{QCD}(q^2,q^2)$ is completely out of control
in perturbative QCD.
A more detailed analysis of $\RcA_{long}$
at the quark level has been recently presented 
in \cite{EKP}, where the leading QCD correction has been estimated. 
Nonetheless, also the final result of \cite{EKP} cannot be considered 
fully conclusive since an arbitrary infrared cutoff is introduced in
order
to avoid the dangerous low $q^2$ region. 

As anticipated, our strategy is to use $f^{QCD}(q^2,q^2)$ to fix the 
high $q^2$ behaviour of the low--energy parameterization
(\ref{eq:fdef}).
The simplest requirement  that we can derive from (\ref{eq:fQCD}) 
is that $f(q^2,q^2)$ must vanish for $q^2 \gsim 4 m_c^2$. 
This condition can be implemented in the phenomenological expression
(\ref{eq:Rint}) in two ways: in a weak sense, assuming 
\beq
\Lambda^2 \lsim 4 m_c^2~,
\label{eq:LambdaM}
\eeq
or in a strong one, imposing the ``sum--rule''
\beq
1+2\alpha+\beta =0~.
\label{eq:sumrule}
\eeq
To be conservative we will use only the weak bound in
(\ref{eq:LambdaM}),
the strong one would have been correct only if the low energy 
parameterization (\ref{eq:fdef}) was valid also above the charm
threshold.
A more realistic constraint on $|1+2\alpha+\beta |$ can be obtained 
imposing the matching between (\ref{eq:fdef})  and (\ref{eq:fQCD}) 
for $\Lambda^2_{QCD} \ll q^2 \ll 4 m_c^2$. In this case from the second
line of (\ref{eq:fqcdapp}) we obtain
\beq
|1+2\alpha+\beta | \simeq \dis{14\over 9} |N_F| \simeq 0.3~.
\label{eq:sum2}
\eeq
Interestingly, this result suggests that the sum--rule 
(\ref{eq:sumrule}) is violated only in a mild way below the 
charm threshold. We recall, for comparison, that naive dimensional 
arguments could not exclude values of $|1+2\alpha+\beta |$ one order
of magnitude larger than in (\ref{eq:sum2}). The smallness of 
$|1+2\alpha+\beta |$ is further supported by the leading QCD correction
to $f^{QCD}$. Indeed, as discussed in \cite{VS,EKP}, 
the main effect of this 
correction is an overall multiplicative factor smaller than one.

Combining (\ref{eq:LambdaM}) and (\ref{eq:sum2}), we believe that 
a realistic bound for the last term in (\ref{eq:Rint}) is given by
\beq
|1+2\alpha+\beta | \ln (\Lambda/m_\beta) < 0.4~.
\label{eq:finalsr}
\eeq
We finally note that is not possible to fix 
the absolute sign of $(1+2\alpha+\beta)$ in 
the framework of perturbative QCD. Indeed, 
since we cannot trust the low $q^2$ limit of the 
perturbative calculation, we are not able to 
fix the relative sign between the unnormalized 
form factor ($F^{QCD}(q^2,q^2)$) and the 
$K_L \to \gamma \gamma$ amplitude ($F(0,0)$). Moreover,
the sign of $F(0,0)$ is not clear also in the framework 
of Chiral Perturbation Theory due to
the cancellation of the lowest--order contributions
to $A(K_L \to \gamma \gamma)$ \cite{DEIN,DP}.

\subsection{Determination of $\alpha$ and $\beta$ in the FMV model}
A more precise, but also more model--dependent, 
determination of  $\alpha$ and $\beta$ can be achieved within 
specific hadronization models. The 
Factorization Model in the Vector couplings (FMV)
was proposed in \cite{DP} 
as a framework to compute the factorizable contributions
to weak vertices involving vector mesons. This model was proven to be
efficient in achieving a satisfactory joint description of the vector 
meson exchange
contributions to $K \rightarrow \pi \gamma \gamma$ and $K_L \rightarrow
\gamma \gamma^*$, giving \cite{DP}
\begin{equation}
\alpha_{FMV} = - \, \frac{256 \pi}{9 \sqrt{2}} \, 
\frac{G_8 \alpha_{em} F_{\pi}}{|F(0,0)|} \, ( 2 h_V + \ell_V) f_V \eta
\, \simeq \, -1.22
\label{eq:alphaoc}
\end{equation}
quite close to the experimental result in 
(\ref{eq:alphaval})
\footnote{~In \cite{DP}
it was concluded that the proper estimate of
the $K_L \rightarrow \gamma \ell^+ \ell^-$ slope should be obtained 
adding to the FMV prediction a contribution generated by a  
weak Vector--Vector transition (the main ingredient of the BMS model)
since they are independent contributions.  
However the two terms have a different momentum 
structure and the BMS one is negligible at
large $q^2$, i.e. in the region where we are interested 
in the value of $(1 + 2 \alpha + \beta)$. }.
In (\ref{eq:alphaoc}) $\ell_V =  3 f_V m_V^2 /
(16 \sqrt{2} \pi^2 F_{\pi}^2)$, 
$f_V$ is fixed from $\Gamma(\rho^0 \rightarrow e^+ e^-)$
to be $|f_V| \simeq 0.20$, $\Gamma(\omega \rightarrow \pi^0 \gamma)$
gives $|h_V| \simeq 0.037$ and $m_V = m_{\rho}$. Moreover,
$G_8 \simeq 9.2 \times 10^{-6} \, \mbox{GeV}^{-2}$ is the effective coupling
of the octet ${\cal O}(p^2)$ weak chiral Lagrangian determined from
$K \rightarrow \pi \pi$ and $\eta \simeq 0.21$ was fixed in 
\cite{DP} from the weak $VP\gamma$ vertex. The application of this model 
to the construction of the
$K_L \rightarrow \gamma^* \gamma^*$ vertex through vector meson
dominance and, consequently, the
Pseudoscalar--Vector--Vector (PVV) weak vertex is straightforward and
gives (assuming only octet contributions)
\begin{equation}
\beta_{FMV} \, = \, \frac{256 \pi}{3 \sqrt{2}} \, 
\frac{G_8 \alpha_{em} m_V^2}{F_{\pi} |F(0,0)|} \, \, f_V^3 h_V \eta
\, \simeq \, 1.43~. 
\label{eq:betaoc}
\end{equation}
Note that the experimental value of the 
$\pi^0 \rightarrow \gamma \gamma^*$ slope implies 
$f_V h_V >0$, thus the relative 
signs of $\alpha_{FMV}$, $\beta_{FMV}$ and $A(K_L \rightarrow \gamma
\gamma)$
are completely fixed. The  overall arbitrary sign is constrained 
by the experimental data on $\alpha$ (i.e.
imposing $\alpha_{FMV}$ to be negative).

Combining the predictions of $\alpha$ and $\beta$ in the FMV model  we
get
\begin{equation}
1 \, + \, 2 \alpha_{FMV} \, + \, \beta_{FMV} \, = \, - 0.01~.
\label{eq:combfmv}
\end{equation}
This result is perfectly consistent with the QCD bound
in (\ref{eq:finalsr}).

\section{Numerical results} 
The theoretical bound on $(1+2\alpha+\beta ) \ln (\Lambda/m_\beta)$
in (\ref{eq:finalsr}), together with 
the experimental determination of $\alpha$ in (\ref{eq:alphaval}),
let us estimate $|\RcA_{long}|$ by means of (\ref{eq:ral11})
and (\ref{eq:Rint}). In order to combine the two pieces of
information we must assume a statistical distribution for 
$(1+2\alpha+\beta ) \ln (\Lambda/m_\beta)$. Assuming for the 
latter a flat distribution between $-0.4$ and $+0.4$,
and combining it with the Gaussian distribution of $\alpha$,
we find
\beq
|\RcA_{long}|< 2.9 \times 10^{-5} \qquad (90\%~\mbox{C.L.})~. 
\label{eq:limRca}
\eeq
The same result is obtained assuming for 
$(1+2\alpha+\beta ) \ln (\Lambda/m_\beta)$ a Gaussian 
distribution with central value $0$ and $\sigma = 0.8/\sqrt{12}$
(the $\sigma$ of the original flat distribution).
However, in this case one can distinguish better 
the various contributions to the limit (\ref{eq:limRca}). 
Indeed we find 
\beqa
|\RcA_{long}| &=& \left[ \frac{2\alpha_{em}^2m_\mu^2\beta_\mu
B(K_L\to\gamma\gamma) }{\pi^2m_K^2} \right]^{1/2} \left|~ 5.25 + 3.47
\alpha +
3(1+2\alpha+\beta ) \ln \frac{\Lambda}{m_\beta}~ \right| \qquad \no \\
& & \label{eq:num23} \\
&=& 1.61 \times 10^{-5} \times |0.41 \pm 0.76 \pm 0.69| =
|0.66 \pm 1.65| \times  10^{-5}~. \no
\eeqa 
Interestingly enough, the knowledge of the absolute sign 
of the central value of $\RcA_{long}$
(i.e. the relative sign between short and long distance contributions)
is 
not very important at this stage, given the large value of 
the error in (\ref{eq:num23}). Moreover, the above expression shows
that at present the largest source of uncertainty is  
generated by the experimental error on $\alpha$.

\medskip
Having derived a numerical estimate for $|\RcA_{long}|$
we are finally able to extract some
short distance information from the measured value of
$B(K_L \to \mu^+ \mu^-)$: 
\begin{enumerate}
\item{\underline{Bound on $\bar\rho$.} 
Using the Bayesian prescription of the Particle Data Group \cite{PDG},
we construct a statistical distribution for $|\RcA_{exp}|$
that eliminates the unphysical values. 
Then, combining it with the 
Gaussian distribution of $\RcA_{long}$ discussed above, 
we obtain a distribution function for 
$(|\RcA_{exp}| + |\RcA_{long}|)$. Finally, using this distribution 
in (\ref{eq:rbound}), together with
${\overline m_t}(m_t) = (167 \pm 6)$~GeV and $|V_{cb}|=(0.040 \pm
0.003)$
\cite{Buras}, we find
\beq
\bar\rho > - 0.38  \qquad \mbox{or} \qquad
\rho     > - 0.42  \qquad (90\%~\mbox{C.L.})~.
\label{eq:fbr}
\eeq
This result is consistent with the recent analysis of
the CKM matrix presented in \cite{Buras}. Unfortunately 
the combined constraints on $\rho$ coming from $\epsilon_K$,
$\Delta m_{B_d}$ and $|V_{ub}|$ already indicate
$\rho \gsim - 0.4$. However, it must be 
stressed that the bound in (\ref{eq:fbr}) is
a $90\%~\mbox{C.L.}$ limit and  it has a different 
statistical significance than the one obtained in \cite{Buras}
scanning on $\pm\sigma$ intervals of the input values.
}
\item{\underline{Bound on $|\Real U_{ds}|$.} 
Similarly to the previous case we can derive a bound on 
$|\Real U_{ds}|$ by means of Eq.~(\ref{eq:ubound}). Treating 
also $|\RcA_{w}|$ as a  statistical variable (assuming 
a flat distribution for $\rho$ between $-1$ and $+1$) 
we find
\beq
 |\Real U_{ds}| < 2.4 \times 10^{-5} \qquad (90\%~\mbox{C.L.})~,
\label{eq:fbu}
\eeq 
confirming the original result of Nir and Silverman \cite{Nir}.
}
\end{enumerate}
\medskip

To conclude, we stress that the bounds in (\ref{eq:fbr})
and (\ref{eq:fbu}) must be considered only as a preliminary result which
can be improved both on the theoretical and experimental sides.
Indeed, as discussed above, the largest source 
of uncertainty in the estimate of $|\RcA_{long}|$ is given  
at present by the error on $\alpha$. Moreover, a 
bound on $\beta$ and/or a measurement of the quadratic 
slope in $K_L\to\gamma \ell^+ \ell^-$ could provide interesting 
consistency checks of our approach. Last, but not least,
a more stringent bound on $|\RcA_{exp}|$ could reduce the error 
in the extraction of short--distance parameters.
Thus a substantial improvement could be foreseen in the near future 
with the advent of new high--precision data in the 
sector of rare $K_L$ decays.

\section*{Acknowledgments}
G.I. wishes to thank G. Buchalla, Y. Grossman and M. Worah
for useful discussions and the hospitality of the Theory Group 
at SLAC, where part of this work was done.

 
\end{document}